\documentclass[pra,preprint,eqsecnum,showpacs]{revtex4-1}
\usepackage{amsmath}
\usepackage{amsfonts}
\usepackage{amssymb}
\usepackage{graphicx}
\usepackage{bm}
\begin{document}
\title{The Salpeter equation and probability current in 
the relativistic Hamiltonian quantum mechanics}
\author{K. Kowalski and J. Rembieli\'nski}
\affiliation{Department of Theoretical Physics, University
of \L\'od\'z, ul.\ Pomorska 149/153, 90-236 \L\'od\'z,
Poland}
\begin{abstract}
The probablity current for a quantum spinless relativistic particle
is introduced based on the Hamiltonian dynamics approach utilizing
the Salpeter equation as an alternative of the Klein-Gordon
equation. The correctness of the presented formalism is illustrated
by examples of exact solutions to the Salpeter equation including
the new ones introduced in this work.
\end{abstract}
\maketitle
\section{Introduction}
The problem which arised in the very early days of quantum mechanics
was finding the relativistic counterpart of the Schr\"odinger
equation.  The most popular choices of the relativistic
quantum mechanical equations for spinless massive particle are the
Klein-Gordon equation \cite{1,2,3,4} and less familiar relativistic 
Schr\"odinger equation which is usually referred to as the spinless Salpeter
equation \cite{5,6,7,8,9,10,11,12,13,14,15,16,17,18,19,20,21,22,23}.   
The latter  can be regarded as the ``square root'' of the Klein-Gordon equation 
and is based on the approach which is sometimes referred to as the relativistic 
Hamiltonian dynamics \cite{24,25}.

The advantage of the Klein-Gordon equation is that 
it is manifestly covariant.  Its well-known flaw related to the fact that
this equation is of second order in the time derivative, is the
problem with the probabilistic interpretation.  Namely the time
component of the probability four-current which in the
nonrelativistic case should coincide with the probability density, can be
negative.  As a matter of fact, as the second order equation the 
Klein-Gordon equation can be cast into system of first order 
equations which can be represented in matrix-form equation.  An example is
the Kemmer equation \cite{26,27} and the two-dimensional system
discussed by Feshbach and Villars \cite{28}.  Nevertheless, the
reduction of this kind does not solve problems with the Klein-Gordon 
equation and generate new ones. Furthermore, the appearance of both signs 
of energies for the solutions to the Klein-Gordon equation leads to the 
occurence of the ``Zitterbewegung'' and the Klein paradox.  The ``
Zitterbewegung'' or trembling motion i.e.\ rapidly oscillatory motion whose 
amplitude and period are of order $\hbar/mc$ and $\hbar/mc^2$, respectively, 
is the result of the interference between positive and negative energy
states.  The Klein paradox relies on possibility of transmission to
states with negative kinetic energy in the electrostatic step-like
potentials.  To circumvent the problem one usually suggests the
creation of the particle-antiparticle pairs, that is the need is
indicated for second quantization as with the Pauli-Weisskopf approach
\cite{29} where the probability current is reinterpreted as a charge
current.  Finally, the scalar product in the space of solutions of the
Klein-Gordon equation, although Lorentz covariant, is not positive
definite which causes serious problems.  Therefore, the severe 
difficulties arise with the physical interpretation 
of the Klein-Gordon equation and it is widely believed that based on this
equation one cannot construct consistent one particle relativistic
quantum mechanics.  

The disadvantage of the Salpeter equation is 
that it is not manifestly covariant.  Another problem which is sometimes 
indicated is the nonlocality of the relativistic Hamiltonian which is the
pseudodifferential operator.  However, as was pointed out by L\"ammerzahl
\cite{22} the nonlocality of the Salpeter equation does not
disturb the light cone structure.  The macrocausality of the second
quantized version of this theory was also reported therein.  In addition,
it has been demonstrated by Foldy \cite{6} that the space 
$L^2({\mathbb R}^3,d^3{\bm x})$ of solutions to the Salpeter equation
is invariant under the Lorentz group transformations which is a direct
consequence of a unitary relationship between $L^2({\mathbb R}^3,d^3{\bm x})$
and the space of the unitary irreducible representation of the inhomogeneous
Lorentz group \cite{25}.  On the other hand, it is clear that nonlocal pseudodifferential 
equations like the Salpeter equation are much more complicated than the 
local differential ones like the Klein-Gordon equation.  In spite of 
difficulties, the great advantage of the Salpeter equation is that 
it possesses solutions of positive energies only, so we have no problems with
paradoxes mentioned above occurring in the case of the Klein-Gordon
equation.  We also point out that agreement of predictions of the spinless
Salpeter equation with the experimental spectrum of mesonic
atoms is as good as in the case of the Klein-Gordon equation
\cite{30}.  Moreover, the possibility of probablistic interpretation
in the quantum case as well as clear classical physical content of
the Salpeter equation was the motivation for its wide usage in the
phenomenological description of the quark-antiquark-gluon system as
a hadron model \cite{31,32}.  

In this work, after discussion of the relationship of the Salpeter 
equation with the corresponding integro-differential equation, 
we introduce and examine the probablility
current derived from the spinless Salpeter equation.  We show 
that such current has all good properties of its nonrelativistic
counterpart. In particular, our analysis shows that the nonlocality 
of the Salpeter equation does not disturb causal propagation of
particles.  The theory is illustrated by the concrete
examples of exact solutions of the Salpeter equation. 
\section{The Salpeter equation}
This section is devoted to the discussion of the basic facts about
the Salpeter equation.  The Hamiltonian of a 
relativistic classical particle subject to the potential $V({\bm x})$ is
\begin{equation}
H=\sqrt{c^2{\bm p}^2+m^2c^4} + V({\bm x}),
\end{equation}
where $m$ is the mass of the paricle, $c$ is the speed of light and
${\bm x}$ is the three-position.  In relativistic quantum mechanics
the system defined by the Hamiltonian (2.1) is described by the
Salpeter equation of the form
\begin{equation}
{\rm i}\hbar\frac{\partial\phi({\bm x},t)}{\partial
t}=[\sqrt{m^2c^4-\hbar^2c^2\Delta}+V({\bm x})]\phi({\bm x},t),
\end{equation}
where $\Delta\equiv{\bm\nabla^2}$.  The Eq.\ (2.2) is obtained by the
quantization procedure utilizing the Newton-Wigner localization scheme 
\cite{6}. In this scheme we have the standard quantization rule
$\hat{\bm x}\to{\bm x}$, and $\hat{\bm p}\to-{\rm i}\hbar{\bm\nabla}$.
As a consequence, the space of solutions of the Salpeter 
equation is the Hilbert space $L^2({\mathbb R}^3,d^3{\bm x})$ with the 
scalar product 
\begin{equation}
\langle \phi|\psi\rangle = \int d^3{\bm x}\,\phi^*({\bm x})\psi({\bm x}).
\end{equation}
Therefore, according to the quantum-mechanical spirit, we should 
identify $|\phi({\bm x},t)|^2$ with the probability 
density $\rho({\bm x},t)$ satisfying the normalization condition:
\begin{equation}
\int d^3{\bm x}\,\rho({\bm x},t)=1.
\end{equation}
Motivated by usage of limiting
procedures in the nonrelativistic and ultrarelativistic case as well
as some dimensional considerations we keep in this section and the 
following one the physical constants $\hbar$ and $c$.  The natural
units $\hbar=1$ and $c=1$ are utilized in Sec.\ IV.  Performing the
Fourier transformation
\begin{equation}
\phi({\bm x},t)=\frac{1}{(2\pi)^\frac{3}{2}\hbar^3}\int d^3{\bm
p}\,e^{{\rm i}\frac{{\bm p}\mbox{\boldmath$\scriptstyle{\cdot}$}
{\bm x}}{\hbar}}\tilde\phi({\bm p},t),
\end{equation}
(in the following, whenever it is clear from the context, we omit
designation of a region of integration) we obtain from (2.2) the 
following equation
\begin{equation}
{\rm i}\hbar\frac{\partial\tilde\phi({\bm p},t)}{\partial t}=[\sqrt
{m^2c^4+{\bm p}^2c^2} + V({\rm i}\hbar{\bm\nabla}_{\bm p})]\tilde\phi({\bm p},t).
\end{equation}
It is clear that (2.6) is the partial differential equation of
finite order only for $V({\bm x})$ polynomial in ${\bm x}$.
Note that $\sqrt{m^2c^4+{\bm p}^2c^2}$ is the so called ``symbol" of
the psudodifferential operator $\sqrt{m^2c^4-\hbar^2c^2\Delta}$, that
is
\begin{equation}
\sqrt{m^2c^4-\hbar^2c^2\Delta}\,\phi({\bm x},t) = \frac{1}{(2\pi)^\frac{3}{2}
\hbar^3}\int d^3{\bm p}\,\sqrt{m^2c^4+{\bm p}^2c^2}\,e^{{\rm i}\frac{{\bm p}
\mbox{\boldmath$\scriptstyle{\cdot}$}{\bm x}}{\hbar}}\tilde\phi({\bm p},t).
\end{equation}
On taking the inverse Fourier transformation
\begin{equation}
\tilde\phi({\bm p},t)=\frac{1}{(2\pi)^\frac{3}{2}}\int d^3{\bm
y}\,e^{-{\rm i}\frac{{\bm p}\mbox{\boldmath$\scriptstyle{\cdot}$}
{\bm y}}{\hbar}}\phi({\bm y},t),
\end{equation}
and making use of the identity \cite{33}
\begin{equation}
\int_{-\infty}^\infty\sqrt{x^2+a^2}\,e^{{\rm i}px}\,dx =
-\frac{2a}{|p|}K_1(a|p|), 
\end{equation}
where $K_\nu(z)$ is the modified Bessel function (Macdonald
function), as well as the differentiation formula satisfied by the Bessel
functions such that
\begin{equation}
K'_1(z)=\frac{1}{z}K_1(z)-K_2(z),
\end{equation}
we get from (2.7) the following formula on the action of the
pseudodifferential operator $\sqrt{m^2c^4-\hbar^2c^2\Delta}$ in the
coordinate representation:
\begin{equation}
\sqrt{m^2c^4-\hbar^2c^2\Delta}\,\phi({\bm x},t) = 
\int d^3{\bm y}\,K({\bm x}-{\bm y})\phi({\bm y},t),
\end{equation}
where the function $K({\bm x}-{\bm y})$ is
\begin{equation}
K({\bm x}-{\bm y}) =
-\frac{2m^2c^3}{(2\pi)^2\hbar}\frac{K_2(\frac{mc}{\hbar}|{\bm x}-{\bm y}|)}
{|{\bm x}-{\bm y}|^2},
\end{equation}
and $|{\bm a}|$ designates the norm of the vector ${\bm a}$.  Thus,
it turns out that the pseudodifferential operator 
$\sqrt{m^2c^4-\hbar^2c^2\Delta}$ can be defined as the integral
operator with the kernel (2.12).  Consequently, the Salpeter
equation (2.2) takes the form of the integro-differential equation
\begin{equation}
{\rm i}\hbar\frac{\partial\phi({\bm x},t)}{\partial
t}=\int d^3{\bm y}\,K({\bm x}-{\bm y})\phi({\bm y},t)+V({\bm x})\phi({\bm x},t).
\end{equation}
We also remark that the nonlocality of the Salpeter equation
is related only to the kinetic energy term described by the integral operator
from the right-hand side of Eq.\ (2.13) and does not depend on the potential.
Therefore, the nonlocality is not connected with potential forces acting on a 
quantum particle.  Consider now the massless limit $m=0$, when the Salpeter
equation is
\begin{equation}
{\rm i}\hbar\frac{\partial\phi({\bm x},t)}{\partial
t}=[\hbar c\sqrt{-\Delta}+V({\bm x})]\phi({\bm x},t).
\end{equation}
Of course, the corresponding Fourier transform $\tilde\phi({\bm p},t)$
fulfils the massless limit of Eq.\ (2.6), i.e.
\begin{equation}
{\rm i}\hbar\frac{\partial\tilde\phi({\bm p},t)}{\partial t}=[c|{\bm p}|
+ V({\rm i}\hbar{\bm\nabla}_{\bm p})]\tilde\phi({\bm p},t).
\end{equation}
Taking the limit $m\to 0$ of Eq.\ (2.12) and using the asymptotic
formula
\begin{equation}
K_2(z) = \frac{2}{z^2},\qquad z\to 0,
\end{equation}
we find for $m=0$
\begin{equation}
K({\bm x}-{\bm y}) = -\frac{2c\hbar}{\pi^2}\frac{1}{|{\bm x}-{\bm
y}|^4},\qquad (m=0).
\end{equation}
Finally, consider the simplest case of a relativistic massless
particle moving in a line, when the Salpeter equation is
\begin{equation}
{\rm i}\hbar\frac{\partial\phi(x,t)}{\partial t} =
\left[\sqrt{m^2c^4-\hbar^2c^2\frac{\partial^2}{\partial
x^2}}+V(x)\right]\phi(x,t).
\end{equation}
On performing the Fourier transform
\begin{equation}
\phi(x,t) = \frac{1}{\sqrt{2\pi}\hbar}\int_{-\infty}^\infty dp
\,e^{\frac{{\rm i}px}{\hbar}}\tilde\phi(p,t),
\end{equation}
we obtain the counterpart of Eq.\ (2.6)
\begin{equation}
{\rm i}\hbar\frac{\partial\tilde\phi(p,t)}{\partial t}=\left[\sqrt
{m^2c^4+p^2c^2} + V\left({\rm i}\hbar\frac{\partial}{\partial p}
\right)\right]\tilde\phi(p,t).
\end{equation} 
Clearly, the one-dimensional version of (2.7) is
\begin{equation}
\sqrt{m^2c^4-\hbar^2c^2\frac{\partial^2}{\partial
x^2}}\,\phi(x,t) = \frac{1}{\sqrt{2\pi}\hbar}\int_{-\infty}^\infty
dp\,\sqrt{m^2c^4+p^2c^2}\,e^{\frac{{\rm i}px}{\hbar}}\tilde\phi(p,t).
\end{equation}
Taking the inverse Fourier transformation
\begin{equation}
\tilde\phi(p,t) = \frac{1}{\sqrt{2\pi}}\int_{-\infty}^\infty dy
\,e^{\frac{-{\rm i}py}{\hbar}}\phi(y,t),
\end{equation}
and using (2.9) we find
\begin{equation}
\sqrt{m^2c^4-\hbar^2c^2\frac{\partial^2}{\partial
x^2}}\,\phi(x,t) = \int_{-\infty}^\infty dy\,K(x-y)\phi(y,t),
\end{equation}
where the kernel $K(x-y)$ is given by
\begin{equation}
K(x-y) = -\frac{mc^2}{\pi}\frac{1}{|x-y|}K_1\left(\frac{mc}{\hbar}|x-y|\right).
\end{equation}
It follows immediately from (2.24) and the asymptotic formula
\begin{equation}
K_1(z) = \frac{1}{z},\qquad z\to 0
\end{equation}
that the kernel in the massless case is of the form
\begin{equation}
K(x-y) = -\frac{c\hbar}{\pi}\frac{1}{(x-y)^2}.
\end{equation}
The relation (2.26) is also a direct consequence of (2.21) with $m=0$, (2.22),
and the identity \cite{33}
\begin{equation}
\int_{-\infty}^\infty |x|e^{{\rm i}px}\,dx = -\frac{2}{p^2}.
\end{equation}
As far as we are aware, the first example of the exact solution to
the Salpeter equation, referring to the massless free particle
moving in a line, was considered by Rosenstein and Horwitz \cite{11}
whereas the solution for the massive free particle was discussed by
Rosenstein and Usher \cite{12}.
The WKB approximation technique to the Salpeter equation was 
developed in Refs. \cite{8} and \cite{9}.
In a series of papers by Hall, Lucha and Sch\"oberl (see for instance
Ref.\ \cite{15}), the energy bounds for the Salpeter equation were
analyzed.  To our best knowledge, the first example of the nontrivial
exact solution to the Salpeter equation in the case of a
particle in ${\mathbb R}^3$, was our solution to the Salpeter
equation for a relativistic massless harmonic oscillator \cite{23}.  More
precisely, we derived the exact stationary wavefunctions and
corresponding exact spectrum of the energy expressed by means of
zeros of the Airy function.  The correctness of the quantization based on
the massless Salpeter equation was confirmed by the good behavior of the
related probability density and expectation values of observables. 
Recently, the new exact solution to the
Salpeter equation has been reported referring to the free
massive particle on a line, with the Gaussian initial wavefunction
\cite{34}.  This solution is very complicated and it has the form of
infinite power series expansion with coefficients expressed by means
of integrals of special functions.  In Sec.\ IV we introduce other
new examples of exact solutions to the Salpeter equation.  In
particular, we derive the nontrivial solution of this
equation for a massless particle in a linear potential. 
\section{Probability current}
In this section we introduce the probability current for a quantum
spinless relativistic particle and discuss its basic properties.
We first discuss the probability density and the probability 
current for the Klein-Gordon equation such that
\begin{equation}
\left[{\rm i}\hbar\frac{\partial}{\partial t} - V({\bm
x})\right]^2\psi({\bm x},t)=(m^2c^4-\hbar^2c^2\Delta)\psi({\bm x},t),
\end{equation}
where the potential $V({\bm x})$ is introduced in the equation by
means of the vector minimal coupling scheme.  We point out that the
denomination ``square root'' of the Klein-Gordon equation, mentioned
in the introduction is appropriate for the Salpeter equation
only in the case of a free particle.  Indeed, by squaring (2.2) we
obtain
\begin{equation}
\left[{\rm i}\hbar\frac{\partial}{\partial t} - V({\bm
x})\right]^2\phi({\bm x},t)=\left\{m^2c^4-\hbar^2c^2\Delta
+[\sqrt{m^2c^4-\hbar^2c^2\Delta},V({\bm x})]\right\}\phi({\bm x},t).
\end{equation}
We recall that in the case of the Klein-Gordon equation the
probability density is given by (compare Ref.\ \cite{35})
\begin{equation}
\rho_{\rm KG} =  \frac{{\rm
i}\hbar}{2mc^2}\left(\psi^*\frac{\partial\psi}{\partial
t}-\psi\frac{\partial\psi^*}{\partial t}+\frac{2{\rm i}}{\hbar}V|\psi|^2\right).
\end{equation}
Since the Klein-Gordon equation is second order in time derivative,
therefore the initial values of $\psi$ and
$\frac{\partial\psi}{\partial t}$ can be arbitrary.  We conclude that
$\rho_{\rm KG}$ can be either positive or negative and the
problem arises with the probablistic interpretation of the
Klein-Gordon equation.  The expression for $\rho_{\rm KG}$ reduces
to the nonrelativistic form in the nonrelativistic limit
$c\to\infty$.  However, to show this one should first assume the
validity of the Salpeter equation (2.2) or equivalently restrict 
to positive energy solutions of the Klein-Gordon equation only.
The formula on the probability current for a Klein-Gordon particle
is identical with the nonrelativistic one, that is we have
\begin{equation}
{\bm j}_{\rm KG} = -\frac{{\rm
i}\hbar}{2m}(\psi^*{\bm\nabla}\psi-\psi{\bm\nabla}\psi^*).
\end{equation}
We stress that neither the probability density (3.3) nor the
probability current (3.4) has the correct limit $m\to0$.  Furthermore,
it can be checked that the continuity equation implied by the massless
Klein-Gordon equation obtained from (3.1) by setting $m=0$ requires
existence of some universal constant with the dimension of length.
This is yet another disadvantage of the Klein-Gordon equation.

We now return to the Salpeter Eq.\ (2.2).
Proceeding analogously as in the case of the nonrelativistic
Schr\"odinger equation we find
\begin{equation}
\frac{\partial|\phi|^2}{\partial t} + \frac{{\rm
i}}{\hbar}(\phi^*\sqrt{m^2c^4-\hbar^2c^2\Delta}\,\phi - \phi
\sqrt{m^2c^4-\hbar^2c^2\Delta}\,\phi^*)=0.
\end{equation}
Now, the probability density $\rho({\bm x},t)$ expressed in terms of the
Fourier transform $\tilde\phi({\bm p},t)$ can be written in the form
\begin{equation}
\rho({\bm x},t) = |\phi({\bm x},t)|^2=\frac{1}{(2\pi)^3\hbar^6}\int d^3{\bm p}d^3
{\bm k}\,e^{{\rm i}\frac{({\bm k}-{\bm p})\mbox{\boldmath$\scriptstyle{\cdot}$}
{\bm x}}{\hbar}}\tilde\phi^*({\bm p},t)\tilde\phi({\bm k},t).
\end{equation}
On using (3.5), (3.6) and the continuity equation
\begin{equation}
\frac{\partial\rho}{\partial t} + {\bm\nabla}{\bm\cdot}{\bm j} = 0,
\end{equation}
where $\rho({\bm x},t)$ is the probability density and 
${\bm j}({\bm x},t)$ is the probability current,
we arrive at the following formula on the probability current:
\begin{equation}
{\bm j}({\bm x},t) = \frac{c}{(2\pi)^3\hbar^6}\int d^3{\bm p}d^3
{\bm k}\,\frac{{\bm p}+{\bm k}}{\sqrt{m^2c^2+{\bm
p}^2}+\sqrt{m^2c^2+{\bm k}^2}}\,e^{{\rm i}\frac{({\bm k}-{\bm p})
\mbox{\boldmath$\scriptstyle{\cdot}$}{\bm x}}{\hbar}}
\tilde\phi^*({\bm p},t)\tilde\phi({\bm k},t).
\end{equation}
It should be noted that the probability current (3.8) has the correct 
nonrelativistic limit $c\to\infty$.  Namely making use of (2.5) we easily find
\begin{equation}
\lim_{c\to\infty}{\bm j}=-\frac{{\rm
i}\hbar}{2m}(\phi^*{\bm\nabla}\phi-\phi{\bm\nabla}\phi^*).
\end{equation}
Furthermore, it follows immediately from (3.8) that
\begin{equation}
\int{\bm j}({\bm x},t)\,d^3{\bm x} = \langle\phi|\hat{\bm v}\phi\rangle,
\end{equation}
where $\hat{\bm v}$ is the operator of the relativistic velocity
\begin{equation}
\hat{\bm v} = \frac{c\hat{\bm p}}{\hat p_0},
\end{equation}
where $\hat{\bm p}=-{\rm i}\hbar{\bm\nabla}$, and $\hat p_0=E/c=
\sqrt{m^2c^2+{\hat{\bm p}}^2}~=~\sqrt{m^2c^2-\hbar^2\Delta}$.
The formula (3.10) is the relativistic counterpart of the well-known 
nonrelativistic expression describing connection of the integral of 
the probability current and average velocity in the given state.  
We stress that the relation (3.10) is not valid in the case of the 
Klein-Gordon equation. Bearing in mind the critique of the
Salpeter equation based on its nonlocality, it is also
worthwhile to point out that the length of the average velocity
(3.11) related to the discussed probability current via Eq.\ (3.10),
does not exceed the speed of light. Indeed, we have
\begin{equation}
\langle\phi|\hat{\bm v}\phi\rangle^2\le\langle\phi|{\hat{\bm v}}^2
\phi\rangle=c^2\langle\phi|\frac{{\hat{\bm p}}^2}{{\hat{\bm
p}}^2+m^2c^2}\phi\rangle\le c^2,
\end{equation}
where the inequality is saturated at $m=0$.  Yet another good property 
of the probability current (3.8) is the existence of the massless limit.  
In fact, putting $m=0$ in (3.8) we immediately get
\begin{equation}
{\bm j}({\bm x},t) = \frac{c}{(2\pi)^3\hbar^6}\int d^3{\bm p}d^3
{\bm k}\,\frac{{\bm p}+{\bm k}}{|{\bm p}|+|{\bm k}|}e^{{\rm i}\frac{({\bm k}-{\bm p})
\mbox{\boldmath$\scriptstyle{\cdot}$}{\bm x}}{\hbar}}
\tilde\phi^*({\bm p},t)\tilde\phi({\bm k},t),\qquad (m=0).
\end{equation}
The probability current can be expressed in terms of the solution $\phi({\bm
x},t)$ to the Salpeter equation.  Indeed, substituting in
(3.8) the Fourier transform (2.8) and using (2.9), (2.10) and the
identity \cite{33}
\begin{equation}
\int_{-\infty}^\infty\frac{e^{{\rm i}px}}{a^2-x^2}\,dx =
\frac{\pi}{a}\sin a|p|,
\end{equation}
we find
\begin{equation}
{\bm j}({\bm x},t) = \int d^3{\bm y}d^3{\bm z}\,{\bm K}({\bm x};{\bm
y},{\bm z})\phi^*({\bm y},t)\phi({\bm z},t),
\end{equation}
where
\begin{eqnarray}
&&{\bm K}({\bm x};{\bm y},{\bm z})\\\nonumber
&&\quad{} = -\frac{{\rm
i}m^2c^3}{(2\pi)^3\hbar^2}({\bm\nabla}_{\bm y}-{\bm\nabla}_{\bm z})
\left\{\frac{1}{|{\bm y}-{\bm x}||{\bm x}-{\bm z}|}\frac{1}{|{\bm y}-{\bm x}|
+|{\bm x}-{\bm z}|}K_2\left[\frac{mc}{\hbar}(|{\bm y}-{\bm x}|
+|{\bm x}-{\bm z}|)\right]\right\}.
\end{eqnarray}
Hence, making use of the theorem on the gradient \cite{36}
\begin{equation}
\int_V d^3{\bm x}\,{\bm\nabla}\varphi({\bm x}) = \int_S 
d{\bm S}\,\varphi({\bm x}),
\end{equation}
where $S$ is the oriented boundary of the volume $V$, and taking into 
account that the Bessel functions $K_\nu(z)$ approach zero as 
$|z|\to\infty$, we arrive at the relation
\begin{equation}
{\bm j}({\bm x},t) = \int d^3{\bm y}d^3{\bm z}\,K(|{\bm y}-{\bm
x}|,|{\bm x}-{\bm z}|)[\phi^*({\bm y},t){\bm\nabla}_{\bm z}
\phi({\bm z},t)-\phi({\bm z},t){\bm\nabla}_{\bm y}\phi^*({\bm y},t)],
\end{equation}
where
\begin{equation}
K(|{\bm u}|,|{\bm w}|)
= -\frac{{\rm i}m^2c^3}{(2\pi)^3\hbar^2}
\frac{1}{|{\bm u}||{\bm w}|}\frac{1}{|{\bm u}|
+|{\bm w}|}K_2\left[\frac{mc}{\hbar}(|{\bm u}|
+|{\bm w}|)\right].
\end{equation}
The formula (3.18) is remarkable.  In fact, it means that the
relativistic probablility current has the form resembling the 
"smeared" nonrelativistic one.

We now discuss the case $m=0$.  On taking the massless limit 
$m\to 0$ and making use of (2.16) we easily get from (3.16) 
the following formula
\begin{equation}
{\bm K}({\bm x};{\bm y},{\bm z})= -\frac{2{\rm i}c}{(2\pi)^3}
({\bm\nabla}_{\bm y}-{\bm\nabla}_{\bm z})
\left[\frac{1}{|{\bm y}-{\bm x}||{\bm x}-{\bm z}|}\frac{1}{(|{\bm y}-{\bm x}|
+|{\bm x}-{\bm z}|)^3}\right]\quad (m=0).
\end{equation}
The relation (3.20) can be also easily derived from
(3.13), (2.8), (3.14), and the identity \cite{33}
\begin{equation}
\int_{-\infty}^\infty\varepsilon(x)\sin ax\,e^{ipx}\,dx =
\frac{2a}{a^2-p^2},
\end{equation}
where $\varepsilon(x)$ is the sign function.  Proceeding analogously
as with (3.15) we get
\begin{equation}
K(|{\bm u}|,|{\bm w}|) = -\frac{2{\rm i}c}{(2\pi)^3}
\frac{1}{|{\bm u}||{\bm w}|}\frac{1}{(|{\bm u}|+|{\bm w}|)^3}\qquad (m=0).
\end{equation}
On the other hand, the formula (3.22) is an immediate consequence of (3.19)
and the asymptotic formula (2.16).

Now, whenever the initial wavefunction is real then the solutions to
the Salpeter equation (2.2) satisfy $\phi^*({\bm x},t)=\phi({\bm
x},-t)$.  Of course, it is a well-known property of the
Schr\"odinger equation.  An immediate consequence of this relation
and (3.15) is ${\bm j}({\bm x},-t)=-{\bm j}({\bm x},t)$, that is
${\bm j}({\bm x},t)$ is an odd function of time.  Furthermore, it
follows easily from (3.16) that if the wavepacket fulfils
$\phi(-{\bm x},t)~=\phi({\bm x}~,t)$ or $\phi(-{\bm
x},t)~=~-\phi({\bm x}~,t)$, then ${\bm j}(-{\bm x},t)=-{\bm j}({\bm
x},t)$.

We now return to the Salpeter equation (2.2).  Using the continuity 
equation (3.7) and the formal power series expansion of the square root
from the right-hand side of (2.2) we get the following formula on the
probability current:
\begin{equation}
{\bm j} = -\frac{{\rm i}mc^2}{\hbar}\sum_{n=1}^\infty\frac{(2n-3)!!}{(2n)!!}
\left(\frac{\hbar}{mc}\right)^{2n}\sum_{k=0}^{2n-1}(-1)^k{\bm\nabla}^k\phi^*
{\bm\nabla}^{2n-k-1}\phi.
\end{equation}
We point out that (3.8) can be formally obtained from (3.23) and
(2.5).  Nevertheless, the formula on the probability current (3.23) 
is mathematically less sound and technically less convenient than the integral 
representation (3.8).  Furthermore, in opposition to (3.8), the current given 
by (3.23) has no correct limit for $m\to0$.

Finally, let us restrict to the simplest case of a relativistic massless
particle on a line.  Equations (3.6), (3.8) and (3.13) take then the following 
form
\begin{eqnarray}
\rho(x,t) &=& |\phi(x,t)|^2=\frac{1}{2\pi\hbar^2}\int dpdk\,e^{{\rm i}\frac{(k-p)x}{\hbar}}
\tilde\phi^*(p,t)\tilde\phi(k,t),\\
j(x,t) &=& \frac{c}{2\pi\hbar^2}\int dpdk\,\frac{p+k}{\sqrt{m^2c^2+p^2}
+\sqrt{m^2c^2+k^2}}e^{{\rm i}\frac{(k-p)x}{\hbar}}\tilde\phi^*(p,t)
\tilde\phi(k,t),\\
j(x,t) &=& \frac{c}{2\pi\hbar^2}\int dpdk\,\frac{p+k}{|p|+|k|}e^{{\rm
i}\frac{(k-p)x}{\hbar}}\tilde\phi^*(p,t)\tilde\phi(k,t),\qquad (m=0),
\end{eqnarray}
respectively; here we have used the Fourier transformation (2.19).  
Let us focus our attention on Eq.\ (3.25).  Taking into account
(2.22), (2.9), and the identity \cite{33}
\begin{equation}
\int_{-\infty}^\infty\frac{e^{{\rm i}px}}{x+a}\,dx =
i\pi\varepsilon(p)e^{-{\rm i}ap},
\end{equation}
we get from (3.25)
\begin{equation}
j(x,t) = \int dydz\,K(x;y,z)\phi^*(y,t)\phi(z,t),
\end{equation}
where
\begin{equation}
K(x;y,z) =
-\frac{imc^2}{2\pi\hbar}\frac{\varepsilon(x-y)-\varepsilon(x-z)}{|y-z|}
K_1\left(\frac{mc}{\hbar}|y-z|\right).
\end{equation}
Hence, taking the limit $m\to 0$ and using (2.25) we find the
following formula on the function $K(x;y,z)$ for the probability current 
in the massless case (3.26)
\begin{equation}
K(x;y,z) = -\frac{{\rm i}c}{2\pi}\frac{\varepsilon(x-y)-\varepsilon(x-z)}
{(y-z)^2},\qquad (m=0).
\end{equation}

We now discuss the case of the massless particle on a line in a more
detail. Let $\phi_\pm(x,t)$ be solution of the Salpeter equation 
\begin{equation}
{\rm i}\frac{\partial\phi(x,t)}{\partial
t}= c\sqrt{-\frac{\partial^2}{\partial x^2}}\,\phi(x,t)
\end{equation}
for a free massless particle with positive (negative) momentum moving in the line to
the right (left).  Then we can identify
\begin{equation}
\tilde\phi_\pm(p,t) = \theta(\pm p)\tilde\phi(p,t),
\end{equation}
where $\theta(p)$ is the Heaviside step function. By means of the Fourier 
transformation (2.19)  for $m=0$, we conclude that $\phi_+$ and $\phi_-$ 
depends on $x-ct$ and $x+ct$, respectively, that is
\begin{equation}
\phi_+(x,t)=\varphi(x-ct),
\end{equation}
and
\begin{equation}
\phi_-(x,t)=\psi(x+ct).
\end{equation}
Let furthermore $\rho_\pm(x,t)=|\phi_\pm(x,t)|^2$ and $j_\pm(x,t)$ 
designate the probability density and probability current, respectively,
corresponding to $\phi_\pm(x,t)$.  An immediate consequence of (3.24)
and (3.26) is
\begin{equation}
j_\pm=\pm c\rho_\pm.
\end{equation}
We point out that the relation (3.35) is a natural
massless relativistic counterpart of the well known nonrelativistic formula
on the probability curent (see (4.29)) expressed by means of the velocity and 
the probability density.  The continuity equation in the one-dimensional case 
such that
\begin{equation}
\frac{\partial \rho}{\partial t} + \frac{\partial j}{\partial x}=0,
\end{equation}
is satisfied identically.  Of course, the general solution $\phi(x,t)$ to the 
Salpeter equation (3.31) is
\begin{equation}
\phi(x,t)=\frac{1}{\sqrt{2}}[\phi_+(x,t)+\phi_-(x,t)]=
\frac{1}{\sqrt{2}}[\varphi(x-ct)+\psi(x+ct)].
\end{equation}
On the other hand, (3.37) is the well-known general solution of the
one-dimensional Klein-Gordon equation for a massless free particle
of the form
\begin{equation}
\left(\frac{\partial^2}{\partial t^2}-c^2\frac{\partial^2}{\partial x^2}
\right)\phi(x,t)=0.
\end{equation}
Notice finally, that for massless particle moving to the right or left, the 
expectation value $\langle \hat x(t)\rangle$ of the position  operator coincides 
with the classical trajectory of a massless particle on a line. Indeed, we have
\begin{equation}
\langle\hat x(t)\rangle = \langle \phi_\pm|\hat x\phi_\pm\rangle = 
\int_{-\infty}^\infty dx\,x\rho(x\mp ct)=\pm ct
+\langle \hat x(0)\rangle.
\end{equation}
\section{Exact solutions}
In this section we introduce the exact solutions to the
Salpeter equation and discuss the corresponding probability
density and probability current.  We begin with the one-dimensional
cases.
\subsection{Free massless particle on a line}
We first study a relativistic free massless particle moving in a line.
The corresponding Salpeter equation can be written in the form
\begin{equation}
{\rm i}\frac{\partial\phi(x,t)}{\partial
t}=\sqrt{-\frac{\partial^2}{\partial x^2}}\,\phi(x,t),
\end{equation}
where we set $c=1$.  Consider the evolution of the (normalized)
wavepacket
\begin{equation}
\phi(x,0) = \sqrt{\frac{2}{\pi}}\frac{a^\frac{3}{2}}{x^2+a^2},\qquad
a>0.
\end{equation}
This package is referred in Ref.\ \cite{11} as to the
``Lorentzian'' one.  On performing the Fourier transformation
\begin{equation}
\phi(x,t) = \frac{1}{\sqrt{2\pi}}\int dp\, e^{{\rm i}px}\tilde\phi(p,t),
\end{equation}
where we set $\hbar=1$, we get from (4.1) the following equation
\begin{equation}
{\rm i}\frac{\partial\tilde\phi(p,t)}{\partial t} =
|p|\tilde\phi(p,t),
\end{equation}
subject to the initial condition
\begin{equation}
\tilde\phi(p,0) = \sqrt{a}e^{-a|p|},
\end{equation}
where the use was made of the identity
\begin{equation}
\int_{-\infty}^\infty \frac{1}{x^2+a^2}\,e^{{\rm i}px}\,dx=\frac{\pi}{a}e^{-a|p|}.
\end{equation}
The solution to the elementary equation (4.4) with the initial
condition (4.5) is
\begin{equation}
\tilde\phi(p,t) = \sqrt{a}e^{-(a+{\rm i}t)|p|}.
\end{equation}
Hence, using (4.3) we obtain the normalized wavefunction at any time
\begin{equation}
\phi(x,t)=\sqrt{\frac{2a}{\pi}}\frac{a+{\rm i}t}{x^2+(a+{\rm i}t)^2}.
\end{equation}
The solution (4.8) was originally derived in Ref.\ \cite{11}, however
the definition of the probability current suggested therein is different
from ours.  An immediate consequence of (4.8) is the following formula on the
probability density
\begin{equation}
\rho(x,t)=|\phi(x,t)|^2=\frac{2a}{\pi}\frac{a^2+t^2}{(x^2-t^2+a^2)^2+4a^2t^2}.
\end{equation}
The time development of the probability density (4.9) is presented in
Fig.\ 1.  Furthermore, taking into account (3.26) and (3.36) we find after some
calculation
\begin{equation}
j(x,t) = \frac{a}{4\pi
t^2}\ln\frac{(x+t)^2+a^2}{(x-t)^2+a^2}-\frac{ax}{\pi
t}\frac{x^2-3t^2+a^2}{(x^2-t^2+a^2)^2+4a^2t^2}.
\end{equation}
The time evolution of the probability current (4.10) is shown in Fig.\ 2.
We point out that there is no singularity in (4.10) for $t=0$.  Namely, we have
$\lim_{t\to 0}j(x,t)=0$.
\begin{figure*}
\centering
\includegraphics[scale=1]{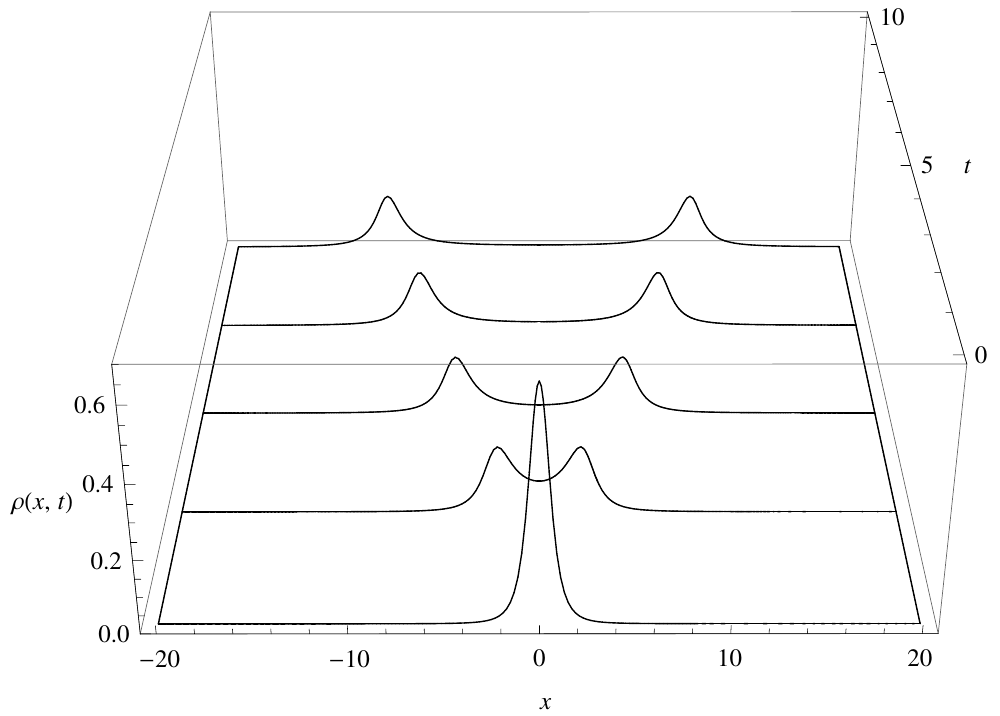}
\caption{The time evolution of the probability density (4.9) related to the
solution of the Salpeter equation for a free massless particle in one
dimension.  Because of choice of the natural units ($c=1$ and $\hbar=1$), the units
of $\rho$, $j$, $x$ and $t$ used in the figures referring to the one-dimensional
case, are ${\rm m}^{-1}$, ${\rm m}^{-1}$, ${\rm m}$ and 
${\rm m}$, respectively.  The parameter $a=1$ m.}
\end{figure*}
\begin{figure*}
\centering
\includegraphics[scale=1]{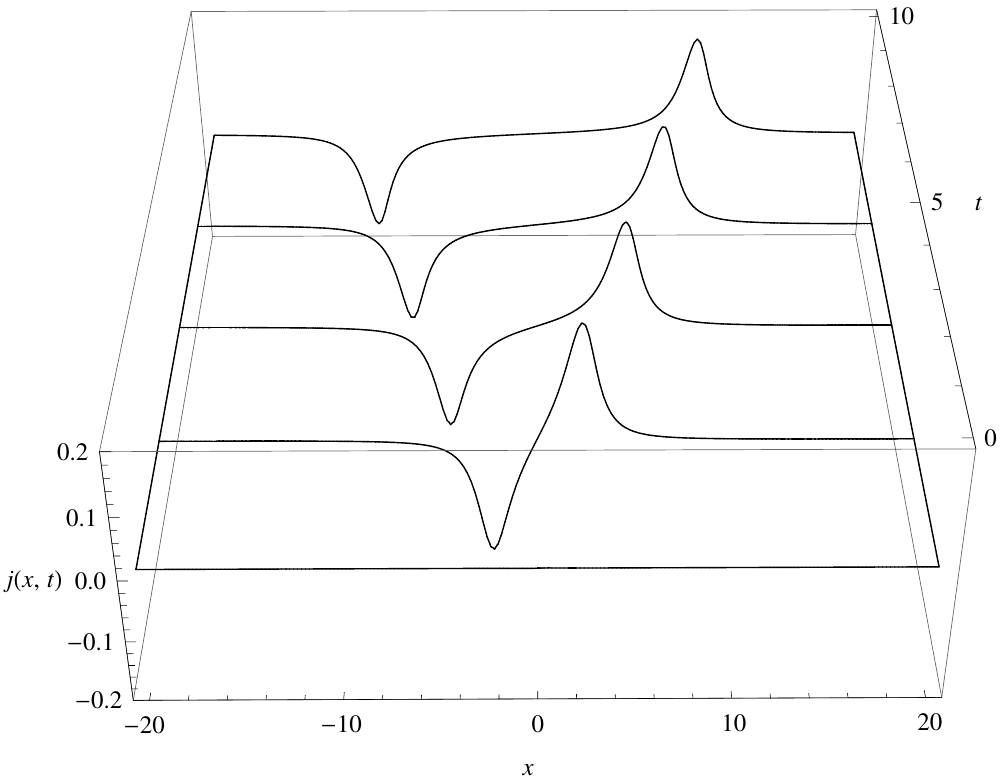}
\caption{The time development of the probability current (4.10) for a free 
massless particle moving in a line.  The parameter $a=1$ m.}
\end{figure*}
\begin{figure*}
\centering
\includegraphics[scale=1]{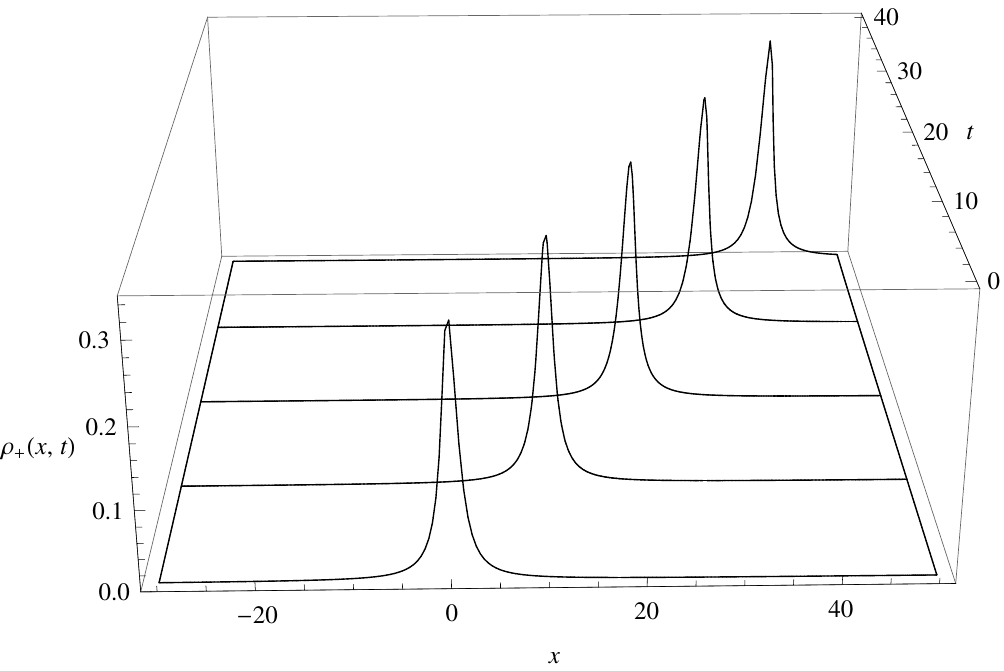}
\caption{The behavior of the probability density $\rho_+(x,t)$
given by (4.13), referring to the case of the free massles particle moving
to the right.  The stable maximum  of the probability density is going with 
the speed of light c=1.}
\end{figure*}

We now return to (4.7).  In view of (3.32) we have
\begin{equation}
\tilde\phi_\pm(p,t)=\sqrt{2a}\theta(\pm p)e^{-(a+{\rm i}t)|p|}.
\end{equation}
Hence, taking into account (4.3) it follows that the wavepacket
referring to the particle moving to the right and left, respectively,
is given by
\begin{equation}
\phi_\pm(x,t) = \sqrt{\frac{a}{\pi}}\frac{\pm{\rm i}}{x\mp t\pm {\rm i}a}.
\end{equation}
Therefore, the corresponding probability density and probability current are
\begin{equation}
\rho_\pm(x,t) = |\phi_\pm(x,t)|^2 = \pm j_\pm(x,t)=
\frac{a}{\pi}\frac{1}{(x\mp t)^2+a^2}.
\end{equation}
The time development of the probability density
$\rho_+(x,t)$ is shown in Fig.\ 3.  
\subsection{Free massive particle on a line}
We now discuss a relativistic free massive particle moving in a
line.  The corresponding Salpeter equation is
\begin{equation}
{\rm i}\frac{\partial\phi(x,t)}{\partial t} =
\sqrt{m^2-\frac{\partial^2}{\partial x^2}}\,\phi(x,t).
\end{equation}
Performing the Fourier transformation we find
\begin{equation}
{\rm i}\frac{\partial\tilde\phi(p,t)}{\partial t}=\sqrt
{m^2+p^2}\,\tilde\phi(p,t).
\end{equation} 
Consider now the particular solution to (4.15) of the form 
\begin{equation}
\tilde\phi(p,t) = \frac{1}{\sqrt{2mK_1(2ma)}}e^{-(a+{\rm i}t)\sqrt{p^2+m^2}}.
\end{equation}
Using the identity \cite{37}
\begin{equation}
\int_0^\infty dx\,\exp(-\alpha\sqrt{x^2+\beta^2})\cos\gamma x =
\frac{\alpha\beta}{\sqrt{\alpha^2+\gamma^2}}K_1(\beta\sqrt{\alpha^2+\gamma^2}),\qquad
{\rm Re}\alpha >0,\, {\rm Re}\beta >0,
\end{equation}
one can easily derive from (4.16) the following (normalized) solution to the Salpeter
equation (4.14):
\begin{equation}
\phi(x,t) = \sqrt{\frac{m}{\pi K_1(2ma)}}\frac{a+{\rm
i}t}{\sqrt{x^2+(a+{\rm i}t)^2}}K_1[m\sqrt{x^2+(a+{\rm i}t)^2}].
\end{equation}
Taking into account the asymptotic formula (2.25)
we find that the solution (4.18) is a generalization of the solution
(4.8) obtained for the massless particle to the case $m>0$.  More
precisely, we have
\begin{equation}
\lim_{m\to 0}\phi(x,t)=\sqrt{\frac{2a}{\pi}}\frac{a+{\rm i}t}{x^2+(a+{\rm i}t)^2}.
\end{equation}
The solution (4.18) was independently derived in Ref.\ \cite{12}.
\begin{figure*}
\centering
\includegraphics[scale=1]{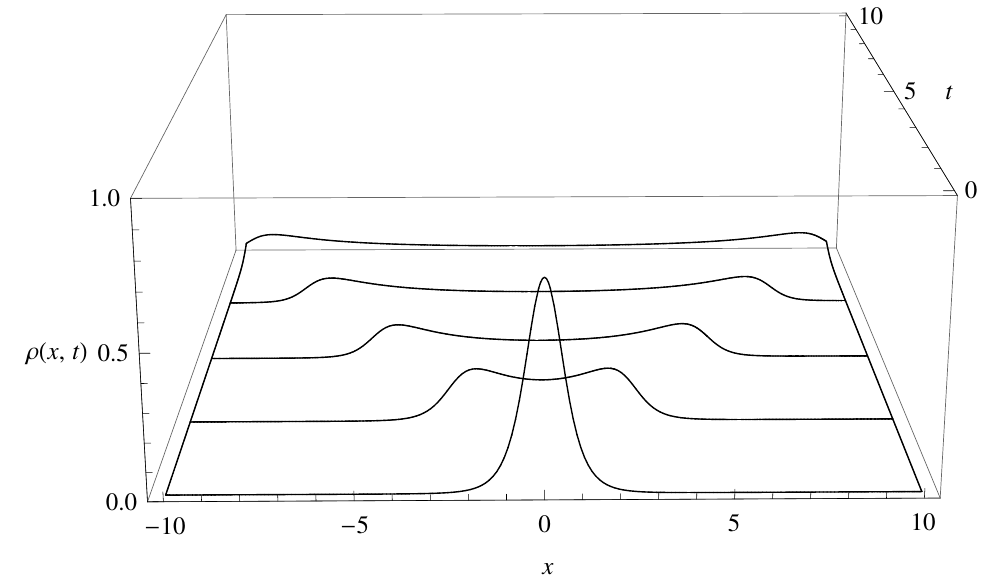}
\caption{The time development of the probability density $\rho(x,t)=|\phi(x,t)|^2$
corresponding to the case of the free massive particle, where $\phi(x,t)$ is given
by (4.18).  The mass $m=0.5$ and $a=1$ m.}
\end{figure*}
The probability density $\rho(x,t)=|\phi(x,t)|^2$ corresponding to the the wavefunction
(4.18) is presented in Fig.\ 4.  It turns out that, in opposition to the massless
case, the wavefunction $\phi(x,t)$ spreads out as time passes.
Finally, taking into account (3.34), (2.10), the identity
\begin{equation}
\frac{1}{z}K_1(z)=-\frac{1}{2}[K_0(z)-K_2(z)],
\end{equation}
and the fact that $\phi(x,t)$ is an even function of $x$ (see
discussion below Eq.\ (3.22)) we find the following formula on the
probability current:
\begin{eqnarray}
j(x,t) =&& \frac{m^2}{\pi K_1(2ma)}\int_0^x dx\,{\rm
Im}\left\{\left[\frac{x^2-(a+{\rm
i}t)^2}{x^2+(a+{\rm i}t)^2}K_2[m\sqrt{x^2+(a+{\rm i}t)^2}]-K_0[m\sqrt{x^2+(a+{\rm i}t)^2}]
\right]\right.\nonumber\\
&&\times\left.\frac{a-{\rm i}t}{\sqrt{x^2+(a-{\rm i}t)^2}}K_1[m\sqrt{x^2+(a-{\rm
i}t)^2}]\right\}.
\end{eqnarray}
The time development of the probablility current (4.21) is shown in Fig.\ 5.
As expected in view of the behavior of wavefunctions, the probablility current 
spreads out.
\begin{figure*}
\centering
\includegraphics[scale=1]{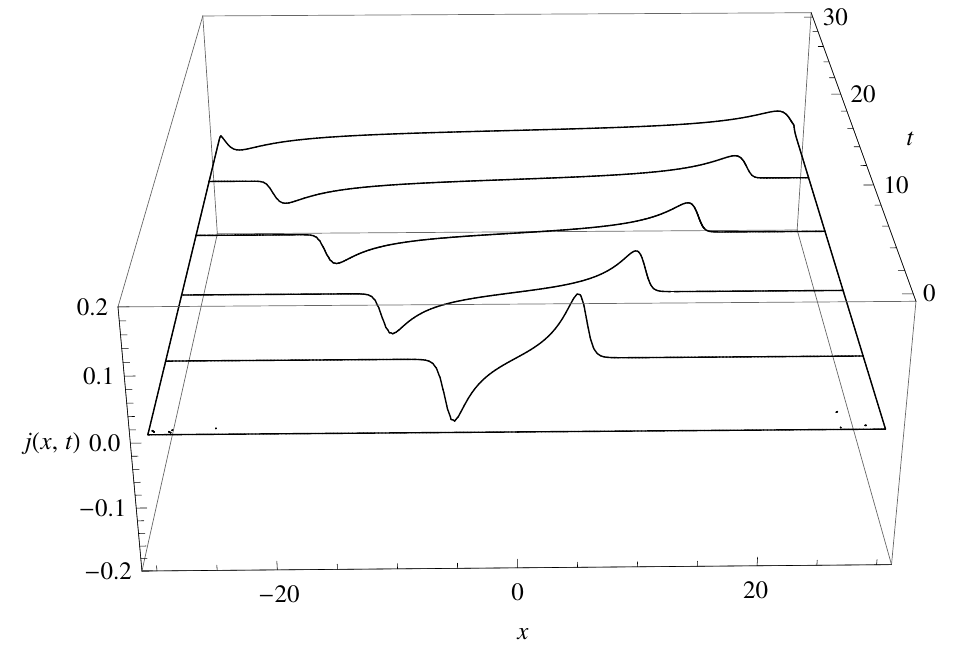}
\caption{The plot of the probability current (4.21) versus time, where $m=0.5$ and 
$a=1$ m.}
\end{figure*}
\subsection{Massless particle in a linear potential}
Our purpose now is to analyze the case of a relativistic massless
particle moving in a line in a linear potential.  Classically, this 
system is defined by the Hamiltonian 
\begin{equation}
H = c|p| +\mu x,
\end{equation}
where $\mu>0$ is a parameter.  The corresponding Hamilton equations lead to the solution 
of the form
\begin{equation}
x(t) = -\frac{c}{\mu}[|-\mu t +p(0)| - p(0)] + x(0). 
\end{equation}
In the following we put $c=1$. Under simplest initial 
conditions  $x(0)=0$, $p(0)=0$ we have $x(t)=-|t|$ (a $\Lambda$-shaped trajectory).
On the quantum level this system is described by the Salpeter equation 
of the form
\begin{equation}
{\rm i}\frac{\partial\phi(x,t)}{\partial
t}=\sqrt{-\frac{\partial^2}{\partial x^2}}\,\phi(x,t)+\mu x\phi(x,t),
\end{equation}
where we set $c=1$ and $\hbar=1$.  The Fourier transform of $\phi(x,t)$
satisfies
\begin{equation}
{\rm i}\frac{\partial\tilde\phi(p,t)}{\partial t} =
|p|\tilde\phi(p,t) + {\rm i}\mu\frac{\partial\tilde\phi(p,t)}{\partial p}.
\end{equation}
The following solution of Eq.\ (4.25) can be derived easily
\begin{equation}
\tilde\phi(p,t) = e^{{\rm i}\frac{\varepsilon(p)p^2}{2\mu}}\chi(p+\mu t),
\end{equation}
where $\chi(p)$ is an arbitrary function.  Now, we choose the initial
wavepacket so that
\begin{equation}
\chi(p) = Ce^{-\frac{\lambda p^2}{2\mu}},
\end{equation}
where $\lambda>0$ is a parameter and $C$ is a normalization constant.
Hence, the normalized wavefunction in the momentum representation is
\begin{equation}
\tilde\phi(p,t) = \left(\frac{\lambda}{\mu\pi}\right)^\frac{1}{4}
e^{{\rm i}\frac{\varepsilon(p)p^2}{2\mu}}e^{-\frac{\lambda (p+\mu t)^2}{2\mu}}.
\end{equation}
Eqs. (4.28) and (4.3) taken together yield the normalized wavefunction such
that
\begin{eqnarray}
&&\phi(x,t) = \frac{1}{2}\left(\frac{\lambda\mu}{\pi}\right)^\frac{1}{4}
e^{-\frac{\lambda\mu}{2}t^2}\left\{\frac{1}{\sqrt{\lambda+{\rm i}}}
e^{\frac{\mu(-\lambda t+{\rm i}x)^2}{2(1+{\rm i})}}
{\rm erfc}\left[\sqrt{\frac{\mu}{2(1+{\rm i})}}(-\lambda t+{\rm i}x)\right]\right.
\\\nonumber
&&\qquad\qquad\left.{}+ \frac{1}{\sqrt{\lambda-{\rm i}}}
e^{\frac{\mu(\lambda t-{\rm i}x)^2}{2(1-{\rm i})}}
{\rm erfc}\left[\sqrt{\frac{\mu}{2(1-{\rm i})}}(\lambda t-{\rm i}x)\right]
\right\},
\end{eqnarray}
where ${\rm erfc}(z) = 1 -\frac{2}{\sqrt{\pi}}\int_0^ze^{-t^2}dt$ is
the complementary error function.  A remarkable property of the corresponding
probability density $\rho(x,t)=|\phi(x,t)|^2$ presented in Fig.\ 6, 
is the behavior of its maxima following the classical $\Lambda$-shaped trajectory.  
Moreover, the expectation value of the position operator, calculated easily in the 
momentum representation is of the form:
\begin{equation}
\langle \hat x(t)\rangle = - \frac{e^{-\lambda\mu t^2}}{\sqrt{\lambda\mu\pi}} - t{\rm erf}
(\sqrt{\lambda\mu} t),
\end{equation}
where ${\rm erf}(t)=\frac{2}{\sqrt{\pi}}\int_0^te^{-\tau^2}\,d\tau$ is the error
function.  Hence, we find the asymptotic formula
\begin{equation}
\langle \hat x(t)\rangle = -|t|,\qquad |t|\gg 1,
\end{equation} 
i.e.\  for $|t|\gg 1$ the average value ot the position operator behaves classically
(see Fig.\ 7).  We also point out that the expectation value of the particle velocity 
$\langle v(t)\rangle= - {\rm erf}(\sqrt{\lambda\mu}t)$ lies between -1 and 1, that is it does not exceed 
the light speed; for $|t| \gg 1$, $|\langle v(t)\rangle|\to 1$, i.e. it reaches 
asymptotically the speed of light.  This is yet another evidence of the correctness of 
our approach based on the Salpeter equation.  The probability current which 
can be derived with the help of (3.36) is too complex to be reproduced herein.  
The time development of the probablility current obtained from computer simulations 
is shown in Fig.\ 8.
\begin{figure*}
\centering
\includegraphics[scale=1]{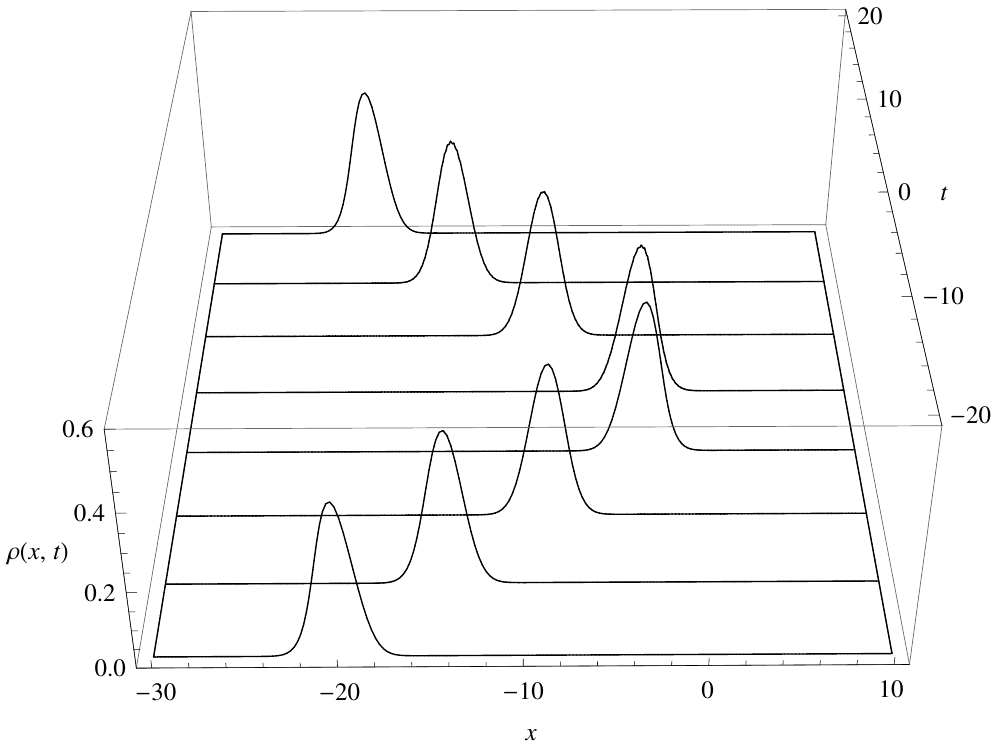}
\caption{The plot of the probability density $\rho(x,t)=|\phi(x,t)|^2$, referring
to the wavefunction (4.29) of the massless particle in a linear potential.
The parameter $\mu=1$ ${\rm m}^{-2}$ and $\lambda=1$.  The classical 
$\Lambda$-shaped dynamics of the maxima of the probability density is easily observed.}
\end{figure*}
\begin{figure*}
\centering
\includegraphics[scale=1]{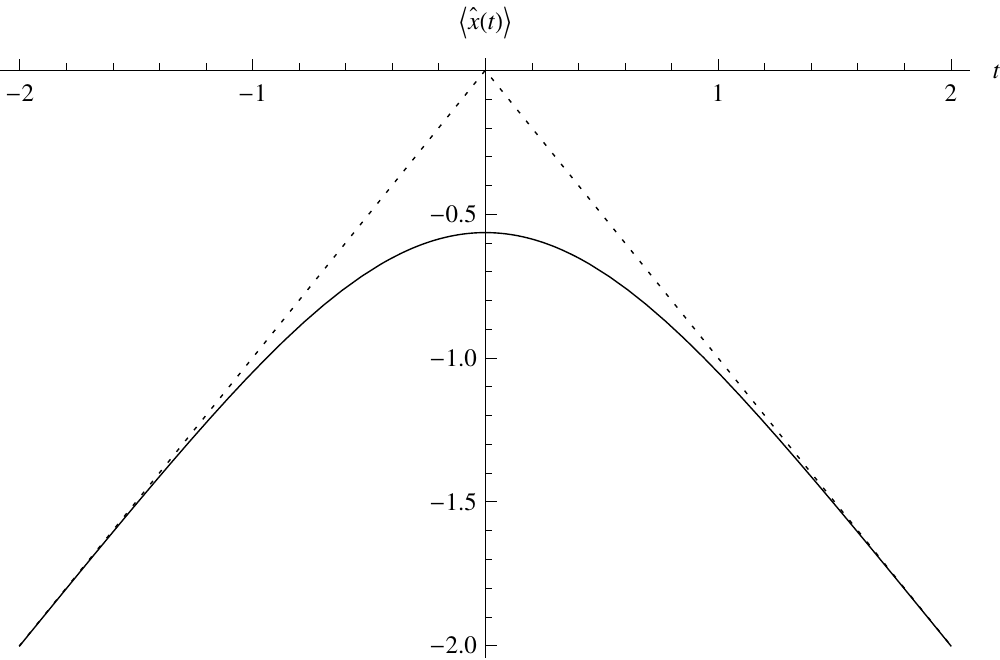}
\caption{The plot of the expectation value of the position operator in the state
(4.29), given by Eq.\ (4.30) with $\mu=1$ ${\rm m}^{-2}$ and $\lambda=1$ versus 
time (solid line).  The dotted line refers to the classical trajectory $x(t)=-|t|$.}
\end{figure*}
\begin{figure*}
\centering
\includegraphics[scale=1]{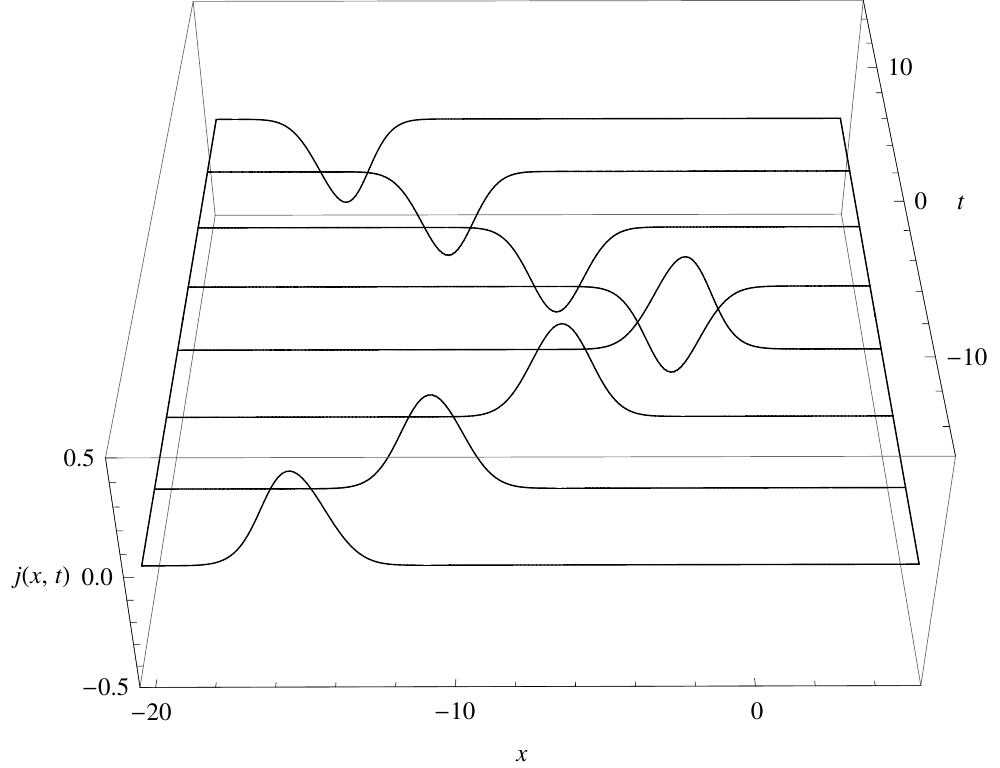}
\caption{The plot of the probability current referring
to the wavefunction (4.29) with $\lambda=1$ versus time.}
\end{figure*}
\subsection{Plane wave solutions}
An easy inspection (compare Ref.\ \cite{22}) shows that the Salpeter
Eq.\ (2.2) with $V=0$, i.e.\ in the case of a free particle, possesses plane 
wave solutions such that
\begin{equation}
\phi({\bm x},t)=Ce^{-\frac{{\rm i}}{\hbar}({\cal E}t-{\bm k}{\bm\cdot}{\bm
x})},
\end{equation}
where ${\cal E}=\sqrt{{m^2c^4+\bm k}^2c^2}$ and $C$ is a normalization
constant.  Taking into account (2.8) and (3.8) or inserting (4.32) into (3.23) 
we obtain the following formula on the corresponding probability current:
\begin{equation}
{\bm j} = \rho{\bm v},
\end{equation}
where $\rho=|\phi|^2=|C|^2$, and ${\bm v}$ is the relativistic 
three-velocity given by
\begin{equation}
{\bm v} = \frac{c^2{\bm k}}{{\cal E}}.
\end{equation}
The relation (4.33) is a natural generalization of the analogous formula
on the probability current which is well-known in the
nonrelativistic quantum mechanics.  We also recall that
nonrelativistic counterpart of (4.33) is a point of departure for the
hydrodynamical formulation of quantum mechanics.  

We point out that in the case of the Klein-Gordon equation the probability 
current for plane waves has the same form as (4.33), that is
\begin{equation}
{\bm j}_{\rm KG}=\rho_{\rm KG}{\bm v}.
\end{equation}
However, the probability density $\rho_{\rm KG}$ such that
\begin{equation}
\rho_{\rm KG}=\frac{|\phi|^2{\cal E}}{mc^2}
\end{equation}
can be either positive or negative depending on the sign of the energy.
\subsection{Massless particle in three dimensions}
In this section we investigate a free massless quantum particle in
three dimensions described by the Salpeter equation
\begin{equation}
{\rm i}\frac{\partial\phi({\bm x},t)}{\partial t} = \sqrt{-\Delta}\,
\phi({\bm x},t),
\end{equation}
where we set $c=1$.  Taking into account the form of the Fourier
transformation (4.7) of the solution to (4.1) corresponding to the
case of a free massless particle in one dimension, one can easily
guess the following solution to (4.37):
\begin{equation}
\phi({\bm x},t) = \frac{C}{(2\pi)^\frac{3}{2}}\int d^3{\bm p}
\,e^{{\rm i}{\bm p}\mbox{\boldmath$\scriptstyle{\cdot}$}
{\bm x}}e^{-it|{\bm p}|}e^{-a|{\bm p}|},
\end{equation}
where $C$ is a normalization constant and $a>0$ is a parameter.  Hence,
we get the normalized solution to (4.37) which is a plausible 
three-dimensional generalization of the solution (4.8), such that
\begin{equation}
\phi({\bm x},t)=\frac{(2a)^\frac{3}{2}}{\pi}\frac{a+{\rm i}t}
{[r^2+(a+{\rm i}t)^2]^2},
\end{equation}
where $r=|{\bm x}|$.  From (4.39) it follows immediately that the 
probability density is
\begin{equation}
\rho({\bm x},t)=|\phi({\bm x},t)|^2=\frac{(2a)^3}{\pi^2}\frac{a^2+t^2}
{[(r^2-t^2+a^2)^2+4a^2t^2]^2}.
\end{equation}
The time evolution of the probability density (4.40)
is presented in Fig.\ 9.  It appears that, in opposition to the case of a 
free particle in one dimesion (see Fig.\ 1), the wavefunction (4.39)
spreads out.
\begin{figure*}
\centering
\includegraphics[scale=1]{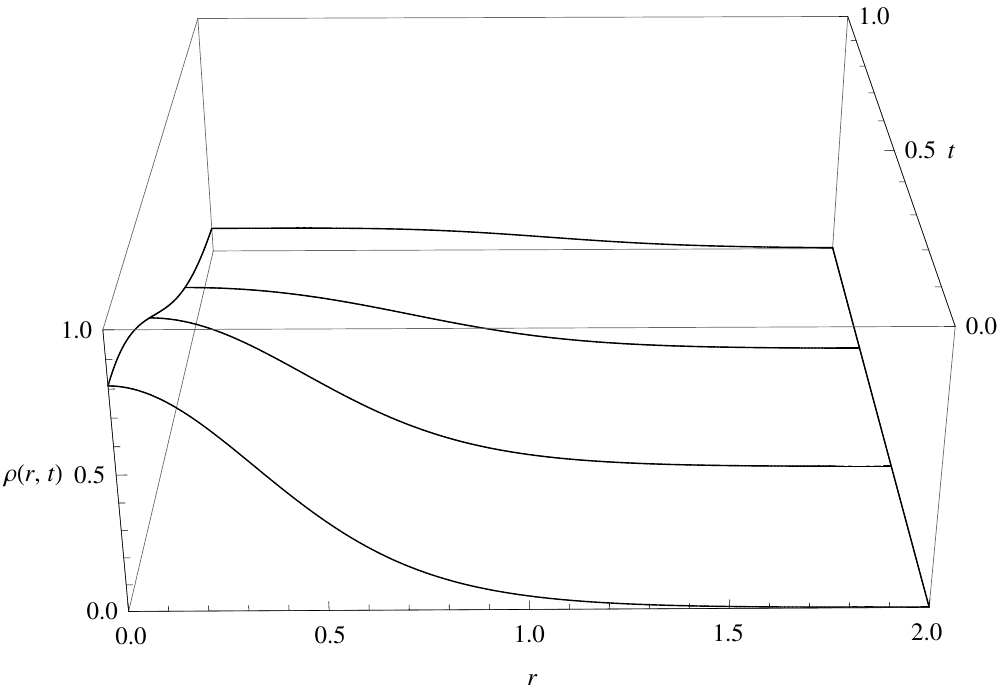}
\caption{The time evolution of the probability density (4.40), where $[\rho]={\rm m}^{-3}$,
$[r]={\rm m}$, and $a=1$ m,
showing the spreading of the wavefunction (4.39).}
\end{figure*}
\begin{figure*}
\centering
\includegraphics[scale=1]{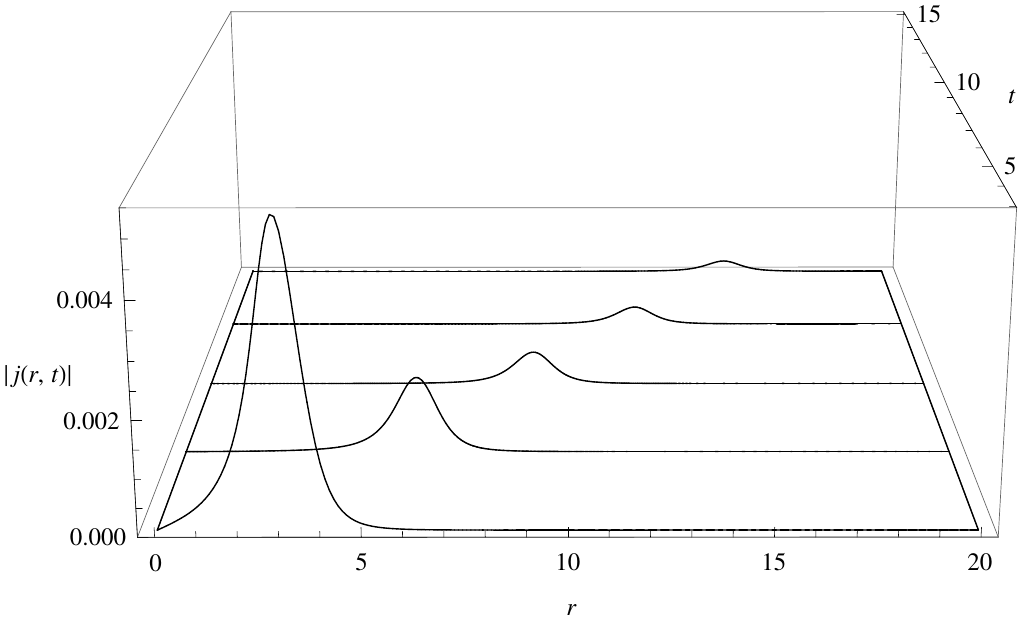}
\caption{The time development of the norm of the probability current (4.41), where
$[j]={\rm m}^{-3}$, and $a=1$ m.}
\end{figure*}
Furthermore, using the definition (3.8) one can derive the following
formula on the probability current:
\begin{eqnarray}
&&{\bm j}({\bm x},t) = \left\{-\frac{a^3}{2\pi^2r^2t^3}
\frac{32r^2t^4+(r^2+t^2+a^2)[3(r^2-t^2+a^2)^2+12a^2t^2-8r^2t^2]}
{[(r^2-t^2+a^2)^2+4a^2t^2]^2}\right.\\\nonumber
&&\qquad\qquad\left.{}+\frac{3a^3}{8\pi^2r^3t^4}\ln\frac{(r+t)^2+a^2}{(r-t)^2+a^2}
\right\}{\bm x}.
\end{eqnarray}
The plot of the norm of the probability current (4.41) is shown in Fig.\ 10.
As expected the norm approaches zero as time passes.
\section{Conclusion}
In this work we study the probability current for a quantum spinless relativistic
particle based on the Salpeter equation as the fundamental equation of the
relativistic quantum mechanics.  The introduced probability current and related
probability density show a very good behavior free of pathologies occuring in the case
of the Klein-Gordon equation such as the negative probability density or the lack of
the massless limit of the probability current.  Referring to the nonlocality which
is often indicated as a grave flaw of the Salpeter equation, we have not
observed any ill behavior of the discussed general as well as particular exact solutions. 
Quite the opposite, the introduced probability current satisfying Eq.\ (3.10) excludes 
in view of (3.12) the particle velocities greater than the speed of light.  We realize that 
the point remains concerning the lack of the manifest covariance of the theory which is usually 
pointed out as the second main disadvantage of the Salpeter equation.  However, it is 
our belief that this flaw can be circumvented by using the preferred frame approach which has 
been already successfully applied for the Lorentz covariant localization in quantum
mechanics \cite{38}, relativistic EPR correlations \cite{39}, and Lorentz covariant 
formulation of classical and quantum statistical mechanics \cite{40,41}.


\begin{references}
\bibitem{1}E. Schr\"odinger, Ann. Phys. {\bf 81}, 109 (1926).
\bibitem{2}O. Klein, Z. Phys. {\bf 37}, 895 (1926).
\bibitem{3}V.A. Fock, Z. Phys. {\bf 38}, 242 (1926); V.A. Fock, 
Z. Phys. {\bf 39}, 226 (1926).
\bibitem{4}W. Gordon, Z. Phys. {\bf 40}, 117 (1926); W. Gordon, 
Z. Phys. {\bf 40}, 121 (1926).
\bibitem{5}E.E. Salpeter, Phys. Rev. {\bf 87}, 328 (1952).
\bibitem{6}L.L. Foldy, Phys. Rev. {\bf 102}, 568 (1956).
\bibitem{7}G. Paiano, Nuovo Cimento {\bf 70}, 339 (1982).
\bibitem{8}P. Cea, P. Calangelo, G. Nardulli, G. Paiano and G.
Preparata, Phys. Rev. D {\bf 26}, 1157 (1982).
\bibitem{9}P. Cea, G. Nardulli and G. Paiano, Phys. Rev. D {\bf 28},
2291 (1983).
\bibitem{10}L.M. Nickisch and L. Durand, Phys. Rev D {\bf 30}, 660
(1984).
\bibitem{11}B. Rosenstein and L.P. Horwitz, J. Phys. A {\bf 18},
2115 (1985).
\bibitem{12}B. Rosenstein and M. Usher, Phys. Rev. D {\bf 36}, 2381 (1987).
\bibitem{13}W. Lucha and F.F. Sch\"oberl, Phys. Rev. D {\bf 50}, 5443
(1994).
\bibitem{14}F. Brau, J. Math. Phys. {\bf 39}, 2254 (1998).
\bibitem{15}R.L. Hall, W. Lucha and F.F. Sch\"oberl, J. Phys. A {\bf 34}, 
5059 (2001),
\bibitem{16}R.L. Hall, W. Lucha and F.F. Sch\"oberl, J. Math. Phys.
{\bf 42}, 5228 (2001).
\bibitem{17}R.L. Hall, W. Lucha and F.F. Sch\"oberl, J. Math. Phys.
{\bf 43}, 5913 (2002).
\bibitem{18}R.L. Hall, W. Lucha and F.F. Sch\"oberl, Int. J. Mod.
Phys. A, {\bf 18}, 2657 (2003).
\bibitem{19}F. Brau, Phys. Lett. A, {\bf 313}, 363 (2003).
\bibitem{20}Zhi-Feng Li, Jin-Jin Liu, W. Lucha and F.F. Sch\"oberl, 
J. Math. Phys. {\bf 46}, 103514 (2005).
\bibitem{21}R.L. Hall and W. Lucha, J. Phys. A {\bf 38}, 7997 (2005).
\bibitem{22}C. L\"ammerzahl, J. Math. Phys. {\bf 34}, 3918 (1993).
\bibitem{23}K. Kowalski and J. Rembieli\'nski, Phys. Rev. A {\bf 81}, 
012118 (2010).
\bibitem{24}F. Gross, {\em Relativistic Quantum Mechanics and Field
Theory} (Wiley, Weinheim, 1993).
\bibitem{25}J. Dimock, {\em Quantum Mechanics and Quantum Field Theory. 
A Mathematical Primer.} (Cambridge University Press, Cambridge, 2011).
\bibitem{26}N. Kemmer, Proc. Roy. Soc. (London) A {\bf 173}, 91 (1931).
\bibitem{27}W.B. Zelezny, Phys. Rev. {\bf 158}, 1223 (1967). 
\bibitem{28}H. Feshbach and F. Villars, Rev. Mod. Phys. {\bf 30}, 24 (1958).
\bibitem{29}W. Pauli and V. Weisskopf, Helv. Phys. Acta {\bf 7}, 709 (1934).
\bibitem{30}J.L. Friar and E.L. Tomusiak, Phys. Rev. C {\bf 29}, 1537 (1984).
\bibitem{31}L.J. Nickisch and L. Durand, Phys. Rev. D {\bf 30}, 660 (1984).
\bibitem{32}J.L. Basdevant and S. Boukraa, Z. Phys. C {\bf 28}, 413 (1985).
\bibitem{33}Yu. A. Brychkov and A.P. Prudnikov, {\em Integral Transformations
of Generalized Functions} (Gordon \& Breach, New York-London, 1989). 
\bibitem{34}D. Babusci, G. Dattoli and M. Quattromini, arXiv:1101.5066 (2011).
\bibitem{35}P. Strange, {\em Relativistic Quantum Mechanics With Applications 
in Condensed Matter and Atomic Physics} (Cambridge University Press, Cambridge, 1998).
\bibitem{36}G.A. Korn and T.M. Korn, {\em Mathematical Handbook} (McGraw-Hill,
New York, 1968).
\bibitem{37}I.S. Gradshteyn and I.M. Ryzhik, {\em Tables of Integrals, Series, 
and Products} (Elsevier, Amsterdam, 2007).
\bibitem{38}P. Caban and J. Rembieli\'nski, Phys. Rev. A {\bf 59}, 4187 (1999).
\bibitem{39}J. Rembieli\'nski and K.A. Smoli\'nski, Phys. Rev. A {\bf 66}, 052114
(2002).
\bibitem{40}K. Kowalski, J. Rembieli\'nski and K.A. Smoli\'nski,
Phys. Rev. D {\bf 76}, 045018 (2007).
\bibitem{41}K. Kowalski, J. Rembieli\'nski and K.A. Smoli\'nski,
Phys. Rev. D {\bf 76}, 127701 (2007).
\end{references}
\end{document}